\documentclass[12pt]{article}

\usepackage{scicite}
\usepackage{graphics}
\usepackage{times}
\usepackage{siunitx}
\usepackage{float}
\usepackage{xr}
\usepackage{bm}
\usepackage{authblk}
\usepackage{afterpage}

\usepackage{graphicx}
\usepackage[labelfont=bf]{caption}
\makeatletter
\newcommand{\settitle}{\@maketitle}
\makeatother
\newenvironment{sciabstract}{%
\begin{quote} \bf}
{\end{quote}}

\newcounter{lastnote}

\title{An ultrastable hard x-ray attosecond split-delay line}

\author{ 
Yanwen Sun$^{1}$, Haoyuan Li$^1$, Yoshio Ichii$^1$, Diling Zhu$^{1\ast}$
\\

\normalsize{\small $^{1}$Linac Coherent Light Source, SLAC National Accelerator Laboratory, Menlo Park, CA 94025, USA} \\

\normalsize{ $^\ast$To whom correspondence should be addressed; E-mail: dlzhu@slac.stanford.edu.}
}

\DeclareUnicodeCharacter{2061}{}
\date{}

\begin{document}
\baselineskip24pt

\maketitle

\begin{sciabstract}

We present a novel split-delay line design for generating hard x-ray attosecond pulse pulse pairs. The design introduces an unconventional delay adjustment mechanism, where an x-ray mirror pair rotation was used for adjusting the path length differential between two beam paths. The exit beam pointing stability is guaranteed  by the mirror-pair self-compensating geometry, therefore enabling stable continuous delay adjustments. We present a parameter study for this concept covering 5-11 keV photon energies with high efficiency over a delay time coverage window of 20-femstosecond with sub-20 attosecond scanning resolution. Wavefront simulations incorporating realistic mirror parameters demonstrate that the system achieves high throughput and is capable of delivering high-peak-intensity pulses.
Attosecond x-ray pump x-ray probe capability enabled by such a delay line is poised to unlock a wide range of hard x-ray nonlinear spectroscopy measurements at sub-femtosecond timescales for the first time.
\end{sciabstract}

\section{Introduction}
The advent of attosecond pulses, produced through high-harmonic generation, has revolutionized ultrafast science by enabling time-resolved studies with unprecedented temporal precision~\cite{krausz2009attosecond}. The attosecond time resolution achieved in experimental techniques, such as pump-probe spectroscopy, is crucial for accessing~\cite{hentschel2001attosecond} and even controlling~\cite{baltuvska2003attosecond} electronic structure and dynamics in matter at their natural timescales—key to understanding quantum phenomena and chemical reactions. Recently, free-electron laser (FEL) facilities have successfully generated attosecond hard x-ray pulses by single spike lasing within a high-current peak of the electron bunch~\cite{marinelli2017experimental, huang2017generating, duris2020tunable, malyzhenkov2020single, yan2024terawatt}. This new mode of operation holds great potential for exploring electron motions with the desired, yet previously inaccessible spatial resolution offered by the atomic-scale wavelengths of x-rays (see Ref~\cite{zhu2024attosecond} and the references therein). 

Pump-probe methodology requires stable and precise temporal control using delay line optics. 
In the hard x-ray range, achieving high system throughput in a delay line requires grazing incidence mirror optics, thus large mirror length to fit the long beam footprints. 
To maintain output beam parallelism, conventional delay lines often employ four reflections per arm~\cite{mandal2021attosecond}. Adapting a similar scheme leads to system sizes of several meters~\cite{roling2020split}. 
Reducing the number of optical elements not only decreases the overall footprint but also improves output beam stability by minimizing vibrational contributions from individual optics. 
Pioneering approaches have explored using just two mirrors for wavefront splitting and delay adjustment via relative angular and translational motions~\cite{murphy2012mirror}.
However, this geometry cannot maintain beam pointing when adjusting delay time, and does not cover time zero.
The second limitation also applies to accelerator-based two-pulse delivery schemes~\cite{duris2020tunable}.
Here, we present a compact ultrastable split-delay line using four broadband mirrors.
This design enables continuous delay scans in a ~20 fs window crossing time zero with fixed output beam angles. 
\section{Optical system design}

\begin{figure*}[h!]
\centerline{\includegraphics[width=0.9\linewidth]{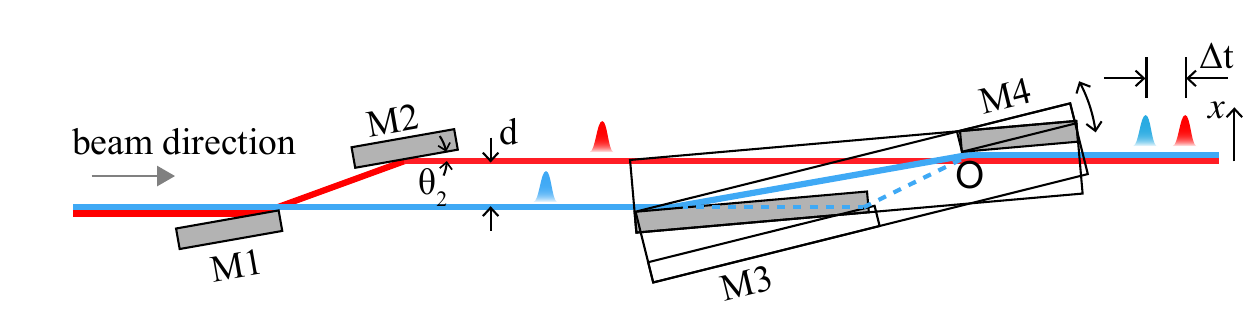}}
\caption{Schematics of the hard x-ray attosecond split-delay line based on 4 mirror optics. Here M denotes mirror. Solid and dashed blue lines show the two extreme trajectories of the delay-adjustable branch, as M3/M4 rotate together to adjust delay. 
}
\label{fig:design}
\end{figure*}

Shown in Fig.~\ref{fig:design} is the optical layout for the delay line. Beam splitting is achieved through horizontal wavefront splitting.
The incident attosecond x-ray FEL pulses are wavefront-split at the downstream edge of mirror 1 (denoted as M1).
The red half of the beam is deflected by the mirror pair M1 and M2, and exits parallel to the incoming beam with a horizontal offset, denoted as $d$.
The blue half of the beam glances over the edge, forming the delay-adjustable branch by reflection via M3 and M4. 

The key optical design element for the delay line is the M3/M4 mirror pair, which are rotated together as a monolithic assembly for delay adjustment.
The rotation center is chosen at the upstream edge of M4 (denoted as point O).
The rotation motion controls the blue beam path length as shown in Fig.~\ref{fig:design}.
Rotating as a single assembly, the output beam angular stability is maintained regardless of the wobble and eccentricity of the rotation motion mechanism.
When choosing the gap size to be $d/(2\cos \theta_1)$, the system is able to reach time zero when $\theta_1 = \theta_3$.
Here $\theta_i$ denotes the reflection angle of mirror $i$.
This facilitates an accurate determination of the time delay of the system by x-ray interferometry when the two pulses overlap within the coherence length~\cite{osaka2017characterization}.  

\begin{figure}[h!]
\centerline{\includegraphics[width=0.7\linewidth]{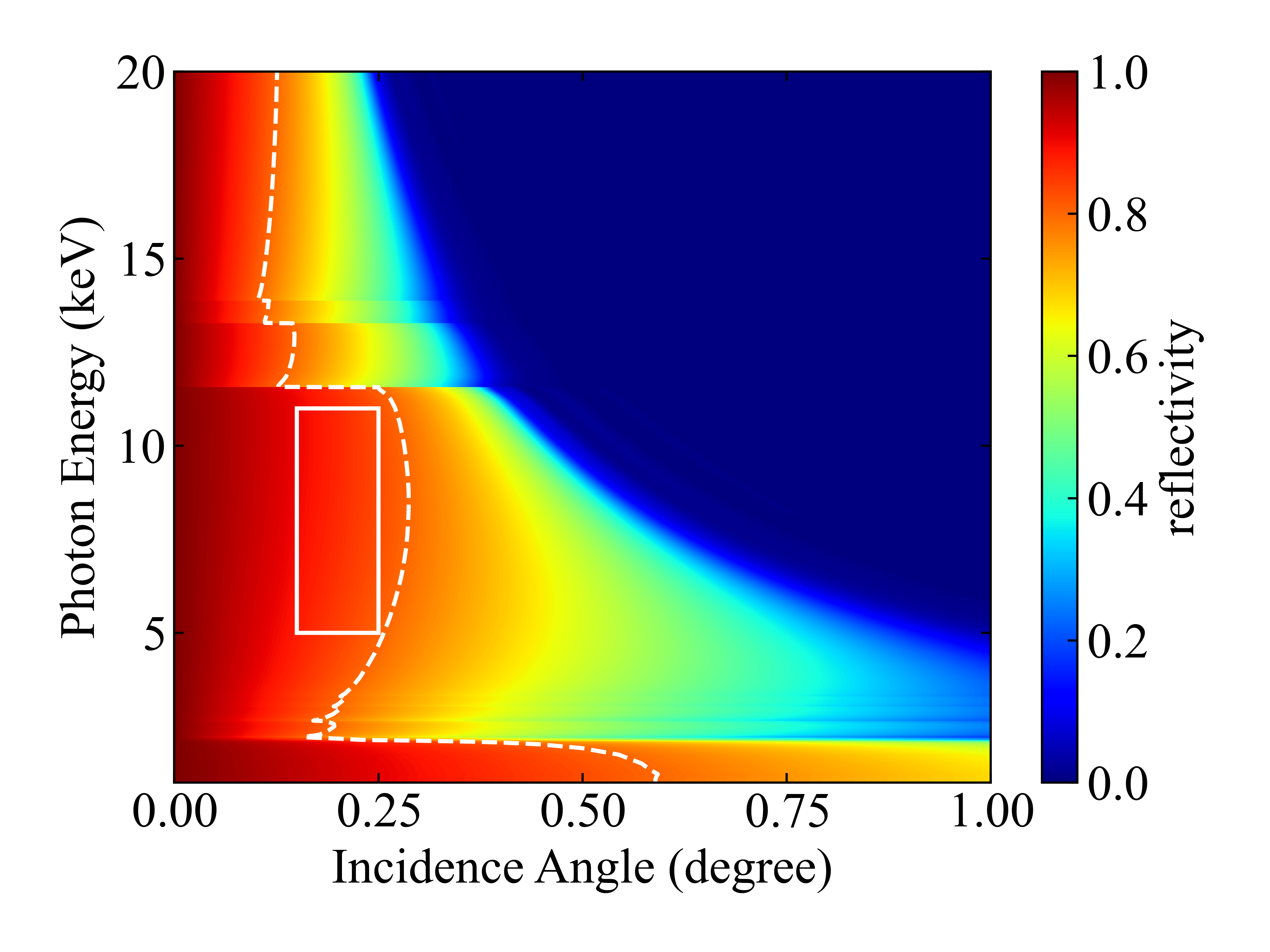}}
\caption{Reflectivity of a pair of mirrors with 30 nm Pt coating on Si substrates. The rectangle outlines the operating region of the delay line, and the dashed white line indicates 80\% reflectivity.}
\label{fig:reflectivity}
\end{figure}

Using $c$ to denote the speedpeed of light, the delay between the two output beams
\begin{equation}
    \Delta t = \frac{d\sin\theta_3}{c\cos\theta_1} - \frac{d\tan\theta_1}{c}.
    \label{equation: Delta_t}
\end{equation}
Accordingly, the resolution for delay adjustment is related to the offset $d$ and also the resolution of the angular motion of M3/M4 $\delta \theta_3$ with
\begin{equation}
    \delta t = \frac{d\cos\theta_3}{c\cos\theta_1}\delta \theta_3.
     \label{equation: delta_t}
\end{equation}
The delay coverage scales proportionally with the offset $d$ and the reflection angle $\theta_3$ ($\sim \sin \theta_3$).
For hard x-ray mirrors, the reflection angle is limited by the total external reflection critical angle. 
Using metal coatings allows for larger reflection angles. 
Figure.~\ref{fig:reflectivity} displays the calculation of mirror-pair reflectivity based on Snell's law and Fresnel equations~\cite{als2011elements}.
With a 30~nm Platinum (Pt) coating on a Si substrate, it enables extending the reflection angles while maintaining high reflectivity.

By selecting the parameters listed in Table~\ref{table:dimension}, the time delay can be adjusted from -8.73~fs to 8.73~fs as $\theta_{3/4}$ is varied between $0.15^{\circ}$ to $0.25^{\circ}$.
At 0.1~mdeg angular resolution when rotating $\theta_{3/4}$, well within the specification of many commercial rotation stages, 17 as step size can be achieved.
As shown in Fig.~\ref{fig:reflectivity}, within the targeted angular range for photon energies between 5-11 keV, outlined by the white rectangle, the system maintains a transmission $>$ 80\%. 
The lengths of M1,2,4 are chosen targeting a clear aperture size of 0.4~mm, matching half the beam size expected 400~m from the source at the Linac Coherent Light Source (LCLS).
As illustrated schematically by the solid and dashed blue lines in Fig.~\ref{fig:design}, rotating the M3/M4 assembly causes the beam to arrive at different locations on M3.
As a result, the length of M3 is chosen to accommodate both the beam footprint and its spatial walk-off.
Maintaining a stable path length corresponding to 1 as timing accuracy requires the mirrors to have a surface flatness of $\pm 17$~nm PV, well within the state-of-the-art mirror fabrication capability~\cite{yamauchi2002figuring}.  

To maximize delay time stability, this delay line design benefits from the reduced number of optics and motion degrees of freedom.
We intend to operate M1/M2 and M3/M4 as two monolithic pairs of optics.
A 100 nrad angular vibration or drift of each mirror along the rotation axis results in a timing jitter or drift of only 1 as.
A 100 nm positional vibration of one mirror along the x-axis induces a delay jitter of less than 2 as.
Considering diffraction-limited focusing optics, an angular shift of 30 nrad for a single mirror results in a beam motion of approximately 10\% of the focused beam size at 8~keV, assuming an input beam size of 0.4 mm after wavefront splitting. 
Mirror translational vibration mentioned above is demagnified by the focusing geometry to be 1 nm, considering a typical demagnification factor of 100.
Vibrations and drifts along other axes have an even smaller impact on both delay and spatial overlap between the two beams due to the small reflection angles.
As M3 and M4 are rotated together to improve stability, the output beam from the delay adjustable branch will have a slight shift along the $x$ direction, given by
\begin{equation}
    \delta x = \frac{d \cos\theta_3}{\cos \theta_1} - d.
\end{equation}
Here $\delta x = 0$ corresponds to the time-zero position when the offsets of the M1/M2 pair and the M3/M4 pair are equal. 
As $\theta_3$ is scanned between $0.15^{\circ}$ and $0.25^{\circ}$ over the full range of delay adjustment, $\Delta s$ changes from 8 nm to -10 nm, which is negligible compared to the transverse beam size of a few hundred $\mathrm{\mu}$m. 

\begin{table}[h!]
    \centering
    \begin{tabular}{|l|l|l|l|l|}
    \hline
    $\theta_{1,2}$ & $\theta_{3,4}$  & M3 length &  M1,2,4 length & d  \\ \hline
     $0.2^{\circ}$ & $0.15^{\circ} - 0.25^{\circ}$  &   300 mm &  150 mm & 3 mm  \\ \hline
    \end{tabular}
    \caption{Design parameters for the attosecond delay line. We use $\theta_i$ to denote beam incident angle on mirror $i$ and d to denote the horizontal offset of the incident and output beam (See Fig.~\ref{fig:design}).}
    \label{table:dimension}
\end{table}

\begin{figure*}[h!]
\centerline{\includegraphics[width=0.95\linewidth]{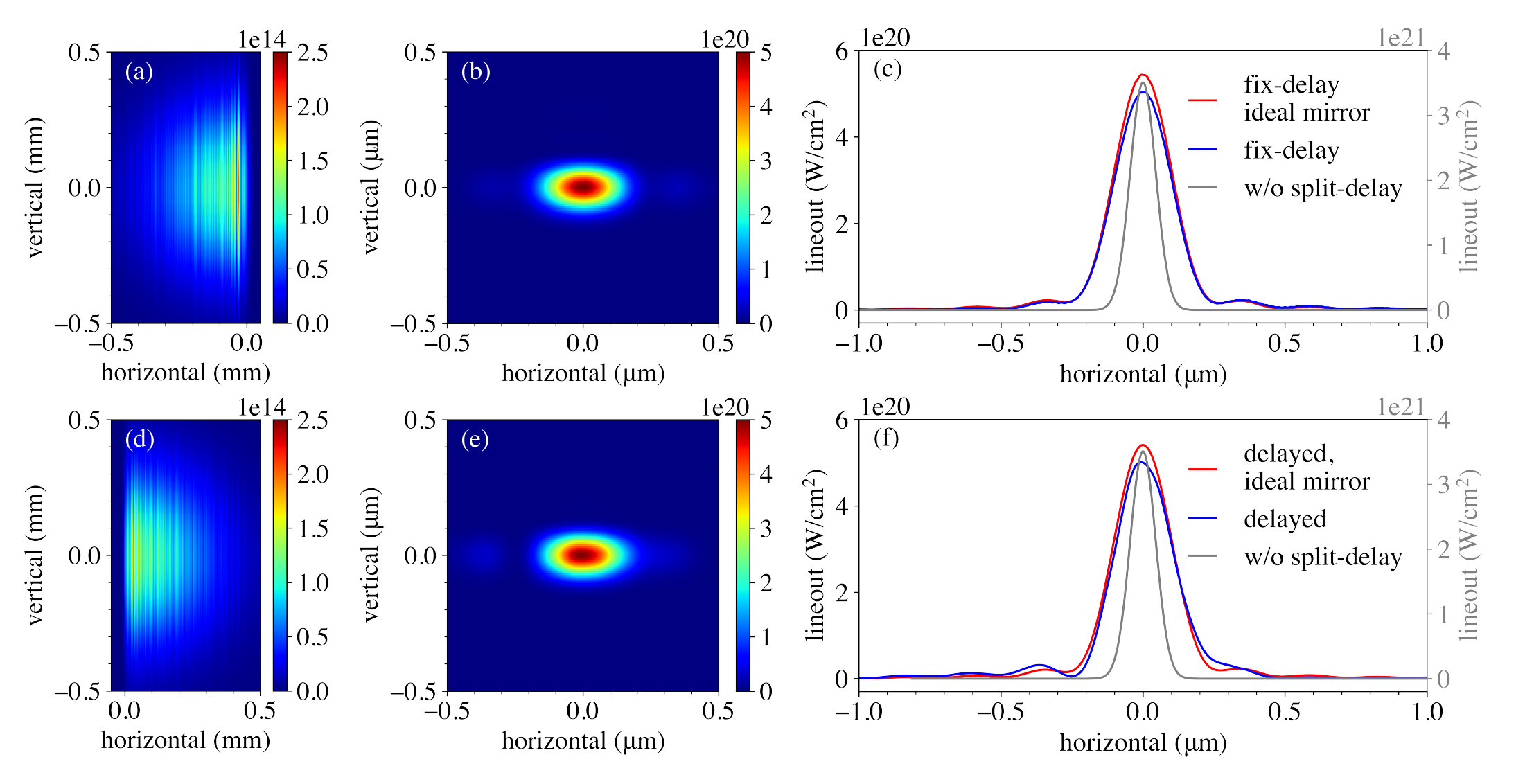}}
\caption{Output pulse intensity profiles for the fix-delay (a, b, c) and delayed branch (d, e, f). The beam profiles at the sample plane, located 2~m downstream of point O in Fig.~\ref{fig:design}, are shown in (a) and (d). To focus the beam at the sample plane, an ideal lens with a focal length of 1 m is placed 1~m downstream of O. The 2D focus profiles are shown in (b) and (d). The horizontal lineouts of the focus are plotted in blue in (c) and (f). They are compared with the focus without the split-delay system (gray), and the case assuming ideal mirrors with zero surface roughness and figure error (red). In (c) and (f), red and blue share the y axis on the left and gray uses the y axis on the right. 
The color bars in all subfigures represent beam peak intensity in units of $\mathrm{W/cm^2}$.}
\label{fig:simulation}
\end{figure*}

\section{Performance analysis}

To evaluate the system performance with realistic mirror parameters, wavefront simulations were performed by propagating a 3D Gaussian beam through the split-delay system using parameters listed in Table~\ref{table:dimension}. 
We utilized SRW (Synchrotron Radiation Workshop) within the OASYS (ORange SYnchrotron Suite) framework~\cite{rebuffi2017oasys,chubar2013wavefront}. A 10~keV beam was modeled with a source size of 18~$\mathrm{\mu}$m (RMS) and a pulse duration of 0.2~fs. 
The split-delay system was positioned 400~m downstream from the source, with an assumed incident pulse energy of 100~$\mathrm{\mu}$J.  
Upon reaching the split-delay system, the beam had a $1/e^2$ diameter of 0.88~mm. 
In the simulation, the beam incidence angles were set to 0.2 degrees for all four mirrors.  
We introduced 0.2~nm (RMS) surface roughness and a shape error of 1~nm (RMS) for each mirror with shape profiles generated based on actual mirror metrology measurements.  
Moreover, to simulate wavefront splitting using a mirror edge with limited polishing quality, we introduced a 50~nm height step extending 2.5 mm from both the beam-splitting and recombining edges on M1 and M4. 

The output beam intensity profiles at the sample plane for the delayed and fix-delay pulses are shown in Fig.~\ref{fig:simulation} (a) and (d), respectively.  
High-frequency vertical streaks are apparent in the beam profiles. 
These patterns arise from edge splitting and the mirror profiles, particularly from figure errors with spatial periods of approximately 1-10 mm~\cite{senba2020improvement}.
The impact of wavefront distortion is directly evaluated by analyzing the beam focus when an ideal lens with a focal length of 1~m is introduced.
The focus profiles are shown in (b) and (e) with their horizontal lineouts plotted in blue in Fig.~\ref{fig:simulation} (c) and (f).
Compared to the ideal mirror case assuming zero roughness and figure error plotted in red, the peak intensity is only a few percent lower.
Note that the specified mirror imperfections subtly modify the focus profiles, particularly for the delayed beam, where the intensity distribution becomes slightly asymmetric, and the side lobe becomes more pronounced.
2D Gaussian fits of the central lobe of the focus yield beam sizes ($1/e^2$ diameter) of 0.391 $\times$ 0.195~${\mathrm{\mu}}m^2$ for the delayed branch and 0.396 × 0.190~${\mathrm{\mu}}m^2$ for the fix-delay branch.
The ideal horizontal focus lineout without the split-delay system is plotted in gray for reference, with a size of 0.180~${\mathrm{\mu}}m$, a factor of 2 smaller due to the full horizontal numerical aperture in the horizontal as expected.
The peak intensity for each branch is approximately $5\times 10^{20}~\mathrm{W/cm^2}$.
This is 15\% of that achieved by direct beam focusing without the split-delay system.
It can be attributed to 3 factors.
First, only half of the intensity is directed into each branch.
Second, the horizontal focus is enlarged by about a factor of 2.
The remaining contributions are from mirror reflectivity and the limited optical length of the mirrors, leading to cutoff of the beam spatial tails.

Finally, we note that the edge splitting impact on the beam wavefront, and the resulting enlarged focus can be mitigated by using spatially separated input beams.
This can be achieved via an upstream beam splitter~\cite{li2021generation}, or utilizing the split-undulator mode with 2 attosecond FEL outputs having different pointings~\cite{guo2024experimental}.

\section{Conclusion}
We have presented the design and performance analysis of a compact hard x-ray attosecond split-delay system layout.
The design utilizes the rotation of a rigid mirror pair to enable continuous delay scans with high output-beam pointing stability.
Having access to a pair of attosecond pulses with tunable delays opens up many new experimental opportunities at the intrinsic time scale of valence electronic motion.
The x-ray pump x-ray probe measurement modality, with time resolution at few and sub-femtosecond regimes, will help us capture charge and energy flow in a previously inaccessible spatial-temporal domain.
Combined with the state-of-the-art nanofocusing capabilities~\cite{yamada2024extreme} to reach high x-ray intensity, systematic studies of nonlinear x-ray matter interaction can be performed without worrying about secondary electronic cascading processes.
The delay line therefore serves as our first step toward systematic time-resolved hard x-ray studies at sub-femtosecond time scale. 

\section*{Acknowledgement}
The authors would like to thank helpful discussions with Agostino Marinelli and Ichiro Inoue. 
\section*{Funding}
 U.S. Department of Energy, Office of Science, Office of Basic Energy Sciences under Contract No. DE-AC02-76SF00515.

\bibliography{sample}
\bibliographystyle{Science}

\end{document}